\documentclass[prl,amssymb,superscriptaddress, twocolumn]{revtex4}

\usepackage{amsmath}
\usepackage{amssymb}
\usepackage{subfigure}
\usepackage{graphicx}
\usepackage{ulem}
\usepackage{multirow}
\usepackage{enumitem}
\usepackage{lipsum, babel}

\newcommand{\Tr}{\textrm{Tr}}

\newcommand{\be}{\begin{equation}}
\newcommand{\ee}{\end{equation}}

\newcommand{\vf}{\varphi}
\newcommand{\lf}{\left}
\newcommand{\rg}{\right}
\newcommand{\ra}{\rangle}
\newcommand{\la}{\langle}

\newcommand{\bea}{\begin{eqnarray}}
\newcommand{\eea}{\end{eqnarray}}

\newcommand{\nn}{\nonumber}

\usepackage{xcolor}
\begin{document}

\title{Superconducting qubit based on twisted cuprate van der Waals heterostructures}

\author{Valentina Brosco}
\thanks{valentina.brosco@cnr.it}
\affiliation{Institute for Complex Systems (ISC) Consiglio Nazionale delle Ricerche and Physics Department University of Rome, ``La Sapienza'', P.le A. Moro, 2 (00185) Roma, Italy}
\author{Giuseppe Serpico}
\affiliation{Max Planck Institute for Chemical Physics of Solids, 01187 Dresden, Germany}
\affiliation{Department of Physics, University of Naples Federico II, Via Cintia, Naples,
80126, Italy}
\author{Valerii Vinokur}
 \affiliation{Terra Quantum AG, Kornhausstrasse 25,
CH-9000 St.\,Gallen, Switzerland}
\affiliation{Physics Department, CUNY, City College of City University of New York, 160 Convent Ave, New York, NY 10031, USA}
\author{Nicola Poccia}
\affiliation{Leibniz Institute for Solid State and Materials Science Dresden (IFW Dresden), 01069 Dresden, Germany}
\author{Uri Vool}
\thanks{uri.vool@cpfs.mpg.de}
\affiliation{Max Planck Institute for Chemical Physics of Solids, 01187 Dresden, Germany}
\date{\today}
\begin{abstract}
\noindent
Van-der-Waals (vdW) assembly enables the fabrication of novel Josephson junctions featuring an atomically sharp interface between two exfoliated and relatively twisted $\rm{Bi_2Sr_2CaCu_2O_{8+x}}$ (Bi2212) flakes.
In a range of twist angles around $45^\circ$, the junction provides a regime where the interlayer two-Cooper pair tunneling dominates the current-phase relation. Here we propose employing this novel junction 
to realize a capacitively shunted qubit that we call\,\textit{flowermon}. The $d$-wave nature of the order parameter endows the flowermon with inherent protection against charge-noise-induced relaxation and quasiparticle-induced dissipation. 
This inherently protected qubit paves the way to a new class of high-coherence hybrid superconducting quantum devices based on unconventional superconductors.
\end{abstract}

\maketitle
Superconducting microwave circuits are macroscopic devices that mimic the quantum properties of atoms. Thanks to their flexible design and strong coupling, superconducting circuits lead the implementation of hardware for quantum technology\,\cite{devoret_2013}.
Currently, the most commonly used superconducting circuits are based on the transmon qubit, having a simple and robust design  with a single Josephson junction  and a shunting capacitor\,\cite{koch2007}. However, the plasmonic nature of  the transmon and the macroscopic size of the shunt capacitor plates reduce its coherence time and enhance cross-talk between qubits, posing significant limitations for the implementation of advanced quantum devices comprising a large number of qubits  and requiring low error rates\,\cite{Brink2018, arute2019}. As a result, a growing number of works explores alternative designs with inherent protection, such as the rhombus\,\cite{blatter_design_2001, doucot_topological_2003,gladchenko_superconducting_2009,bell_protected_2014}, fluxonium\,\cite{Manucharyan2009,nguyen_high-coherence_2019}, bifluxon\cite{kalashnikov_bifluxon_2020}, blochnium\,\cite{pechenezhskiy_superconducting_2020}, KITE\,\cite{smith2020,smith_magnifying_2022}, $0-\pi$\,\cite{brooks_protected_2013,groszkowski_coherence_2018,gyenis_experimental_2021} and nanowire-based\,\cite{larsen_parity-protected_2020,ciaccia_charge-4e_2023} qubits. While significantly higher coherence times are predicted for such circuits in the ideal case, their implementation relies on multiple junctions in a flux loop, making them vulnerable to flux noise and inevitable fabrication imperfections. 
Here we develop a novel single-junction qubit with inherent protection that avoids these drawbacks and provides a robust platform for future quantum devices.
\begin{figure}[t!]
\vspace{-0.5cm}
	\begin{center}
		\includegraphics[width=\columnwidth]{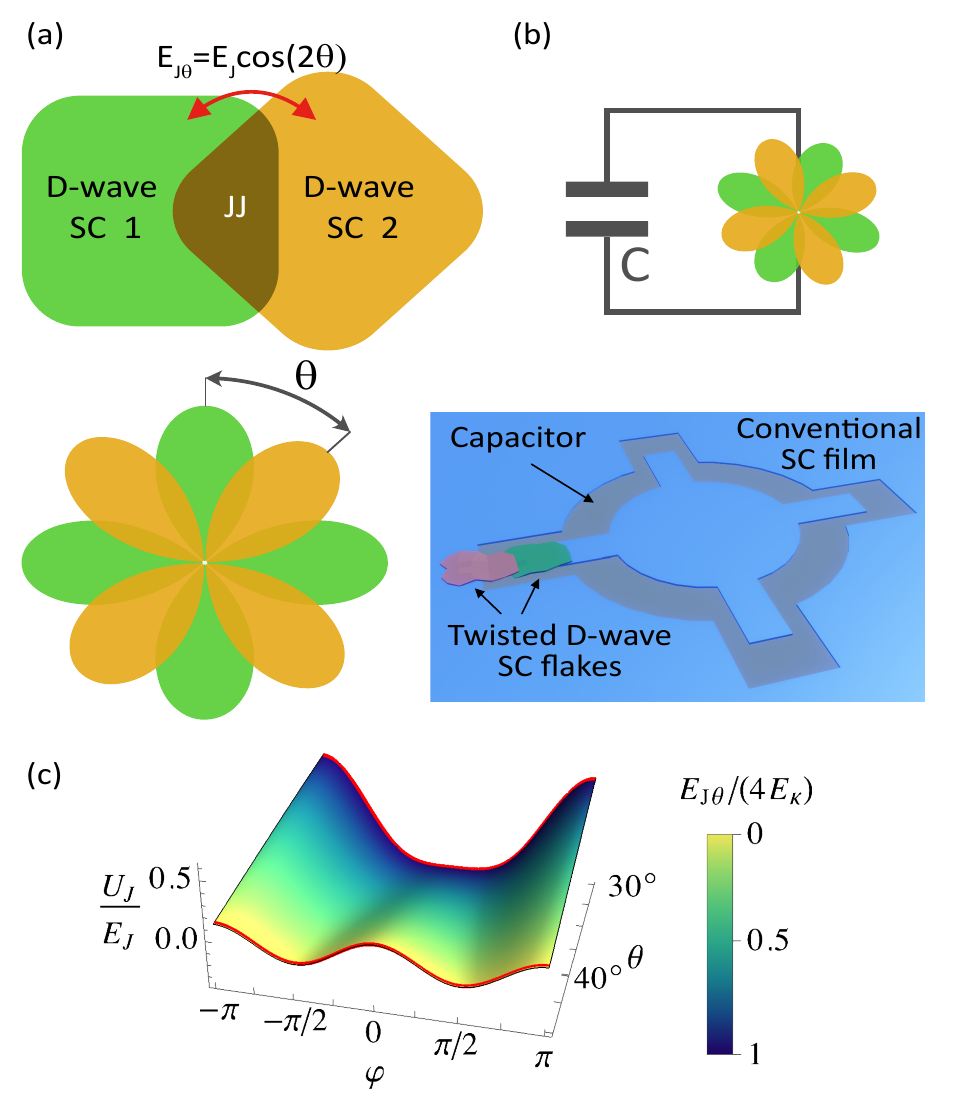}
        \caption{\textbf{Potential design for the flowermon qubit}. \textbf{(a)} A relative twist of two d-wave flakes placed together to form a Josephson junction can suppress Cooper pair tunneling due to momentum mismatch. At $45^\circ$ the mismatch completely suppresses single Cooper pair tunneling, and two-pair tunneling dominates the junction. \textbf{(b)} The design of the flowermon with a single $d$-wave junction shunted by a large capacitor similar to the transmon qubit. The 3D design shows a possible physical implementation with the capacitor pads of a conventional superconductor coupled to the junction.
		\textbf{(c)}\, Josephson's potential for different values of the twist angle.}\label{fig1}
	\end{center}
\end{figure}

Pioneering works\,\cite{Ioffe1999, blais2000, blatter_design_2001,bauch_quantum_2006} proposed to utilize the suppression of tunneling in the $d$-wave based Josephson junctions to realize a qubit with an enhanced coherence. Note that since a single Cooper-pair tunneling between twisted superconducting $d$-wave islands having different orientations is suppressed due to the momentum mismatch, see Fig.\,1a, the contribution of the second-harmonic $\cos(2\varphi)$ of the Josephson energy becomes dominant, and the consequent degeneracy of the spectrum can be exploited for creating a decoherence-protected qubit. In the standard approaches using the grain orientation mismatch, the Josephson junctions at grain boundaries of YBa$_2$Cu$_3$O$_{7-x}$ with relative granule mismatch angle close to $\pi/4$ demonstrated the degeneracy of the ground-state\,\cite{ilichev2001}. 
However, fabrication complexity and the resulting low junction quality hindered the implementation of quantum devices based on grain boundary junctions.

Recent technological progress enabled to preserve nearly perfect superconductivity\,\cite{yu2019high, zhao2019sign} and lattice structure\,\cite{poccia2020spatially} in the isolated atomically thin Bi$_2$Sr$_2$CaCu$_2$O$_{8+x}$ (Bi2212) crystals. This enabled creation of the Bi2212-based junctions, employing a novel technique that allowed to control the diffusion of oxygen interstitials, which is the main source of detrimental disorder in cuprate Bi2212. A cryogenic stacking protocol has been developed utilizing the freezing of the motion of the oxygen interstitials at temperatures below 200\,K\,\cite{poccia2011evolution}, which allowed manufacturing the vdW flakes and creation of an atomically sharp interface between Bi2212 crystals. Such protocol enables the fabrication of high-quality Josephson junctions, displaying a strong dependence of the Josephson energy on the twist angle\,\cite{zhao2021, Lee2023Encapsulating, martini2023twisted}.

Here we propose to utilize such a junction as a platform for a novel highly-protected qubit that we call\,\textit{flowermon}. We focus on the simplest design comprising a single capacitively shunted twisted vdW Josephson junction, as sketched in Fig.\,\ref{fig1}(b). We quantitatively analyze  the behavior of the flowermon qubit, highlighting possible decoherence mechanisms for experimentally-available parameters, and discuss possible control and measurement schemes. 



The flowermon junction has a twist angle $\theta$ between the axis associated with the $d$-wave order parameters of two overlapping superconducting flakes, see Fig. 1a, and is shunted by a capacitor with charging energy $E_{\mathrm C}=e^2/(2C)$ where $C$ is the shunting capacitance which dominates over the internal junction capacitance, see Fig. 1b.
\begin{figure}[t]
	\begin{center}
		\includegraphics[width=\columnwidth]{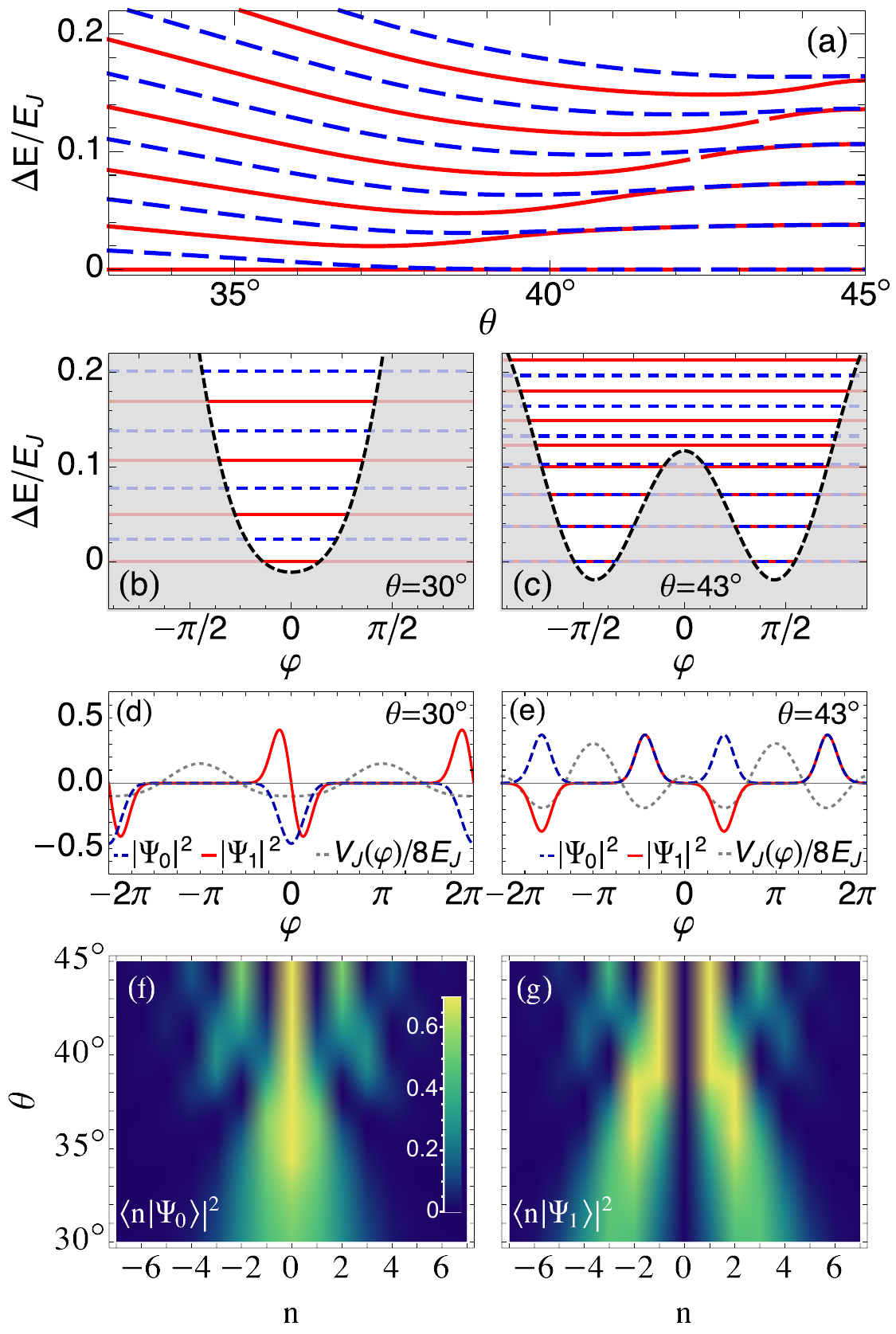}
		\caption{\textbf{Flowermon low-energy spectrum}. \textbf{(a)} Energy levels of the flowermon qubit as a function of the angle $\theta$. Even and odd levels are shown in red and blue lines respectively; as $\theta$ approaches $\pi/4$, the levels combine into quasi-degenerate doublets. Here we assume $E_{\mathrm{\kappa}}/E_{\text J}=0.1$ and $E_J/E_{\text C}=2000$. \textbf{(b)} and \textbf{(c)} Level structure and potential energy for $\theta=30^\circ$ and $\theta=43^\circ$. \textbf{(d)} and \textbf{(e)} The wave-functions for the ground and first excited state $|\Psi_0\ra$ and $|\Psi_1\ra$ in the phase basis for $\theta=30^\circ$ and $\theta=43^\circ$. \textbf{(f)} and \textbf{(g)} Evolution of the structure of $|\Psi_0\ra$ and $|\Psi_1\ra$ in the charge basis as a function of the twisting angle $\theta$.}\label{fig2}
	\end{center}
\end{figure}
The circuit Hamiltonian is
\bea
H= 4E_{\text C} (\hat n-n_{\text g})^2-E_{\text J\theta}\!\cos(\hat \vf)+E_{\mathrm\kappa}\!\cos(2\hat \vf)\label{ham}\,,
\eea
where $\hat \vf$ denotes the phase  difference across the junction and $\hat n$ is the conjugate charge.
 In the above equation $E_{\mathrm J\theta}$ and $E_{\mathrm{\kappa}}$  represent, respectively, the energy associated with the coherent tunneling of single and double Cooper pairs  across the junction, while the gate charge $n_{\text g}$ accounts for charge fluctuations induced by external electric fields.
 The competition between the two Josephson tunneling terms arises due to the peculiar dependence of the Josephson energy on the twisting angle, $E_{\mathrm J\theta}=E_{\mathrm J}\cos(2\theta)$, associated with the $d$-wave structure of the order parameter as has been recently highlighted in Refs.\,\cite{can2021,tummuru2022a} and confirmed experimentally in Ref.\,\cite{zhao2021}.
 The shape of the Josephson potential, $U_{\mathrm J}(\varphi)$ as a function of the twisting angle, is shown in Fig.\,\ref{fig1}(c).
 %
As $\theta$ increases from $0$ to $\pi/4$, the Josephson potential develops a symmetric double-well structure 
with two minima at $\varphi=\pm\varphi_{0}$ with $\varphi_0=\arccos(E_{\mathrm J\theta}/4E_{\mathrm{\kappa}})$ separated by a barrier 
 \be\label{eq:barrier} \Delta U_{\mathrm J}=\left(E_{\mathrm J\theta}-4 E_{\kappa}\right)^2/8 E_{\kappa}.\ee
Using\,Eq.\,\eqref{eq:barrier},\,we\,express\,the\,critical\,angle\,at\,which\,the double-well structure arises as $\theta_{\mathrm c}=\frac{1}{2}\arccos(4E_{\kappa}/E_{\mathrm J})$. 
The ratio $E_{\mathrm \kappa}/E_{\mathrm J}$ thus determines the range of angles in which the qubit can be realized. There is some discrepancy between the theoretical estimates of this ratio\,\cite{can2021} as 0.1-0.2, and experimental observations\,\cite{zhao2021} giving 0.02-0.05. Throughout the paper, we use $E_{\kappa}/E_{\mathrm J}=0.1$; other values will lead to the renormalization of the operating angles. 
Note that $E_{\mathrm J}\propto\Delta$ and $E_{\mathrm \kappa}\propto\Delta^2$, hence high-temperature superconductors such as Bi2212 have a natural advantage due to higher two-Copper-pair tunneling rates. 
Additional effects such as inhomogeneity in the twisted junction can further increase $E_{\mathrm{\kappa}}$\,\cite{yuan2023}. For further discussion of the relation between junction and qubit parameters, see supplementary information\,\cite{supplementary}.

For $\theta_{\mathrm c}<\theta\lesssim\pi/4$ and sufficiently small $E_{\mathrm C}$,  the low-energy spectrum reduces to a set of quasi-degenerate doublets, see Fig.\,\ref{fig2}(a).  
Splitting of each doublet is associated with the tunneling of the phase between the two minima and thus scales exponentially with the ratio $\Delta U_{\mathrm J}/E_{\mathrm C}$\,\cite{smith2020,supplementary}. 
This transition is shown in Fig.\,\ref{fig2}(b),(c) where at $\theta=30^\circ$ the spectrum is quasi-harmonic, reminiscent of the spectrum of the transmon qubit, while at $\theta=43^\circ$ the double-well fully develops, and the lowest energy levels are doubly degenerate. 
In the latter case, the phase is not able to tunnel between two adjacent minima, and, as shown in Fig.\,\ref{fig2}(e), the corresponding wave functions are localized in two energy wells.
The quasi-degeneracy of the levels is associated with the symmetry of the wave functions.
As the angle approaches $\pi/4$ the wave functions of the even and odd energy levels develop a distinct parity in the charge basis.  
Specifically, at high twisting angles, $\theta\gtrsim 40^\circ$, the ground state $|\psi_0\ra$ contains only even Cooper pair number states, while the first excited state $|\psi_1\ra$ contains only odd Cooper pair number states, see Fig.\,\ref{fig2}(f),(g). 
Therefore, when the Josephson energy is completely dominated by double Cooper pair tunneling, the Cooper pair number parity is conserved. 

\begin{figure}[t]
	\begin{center}
		\includegraphics[width=\columnwidth]{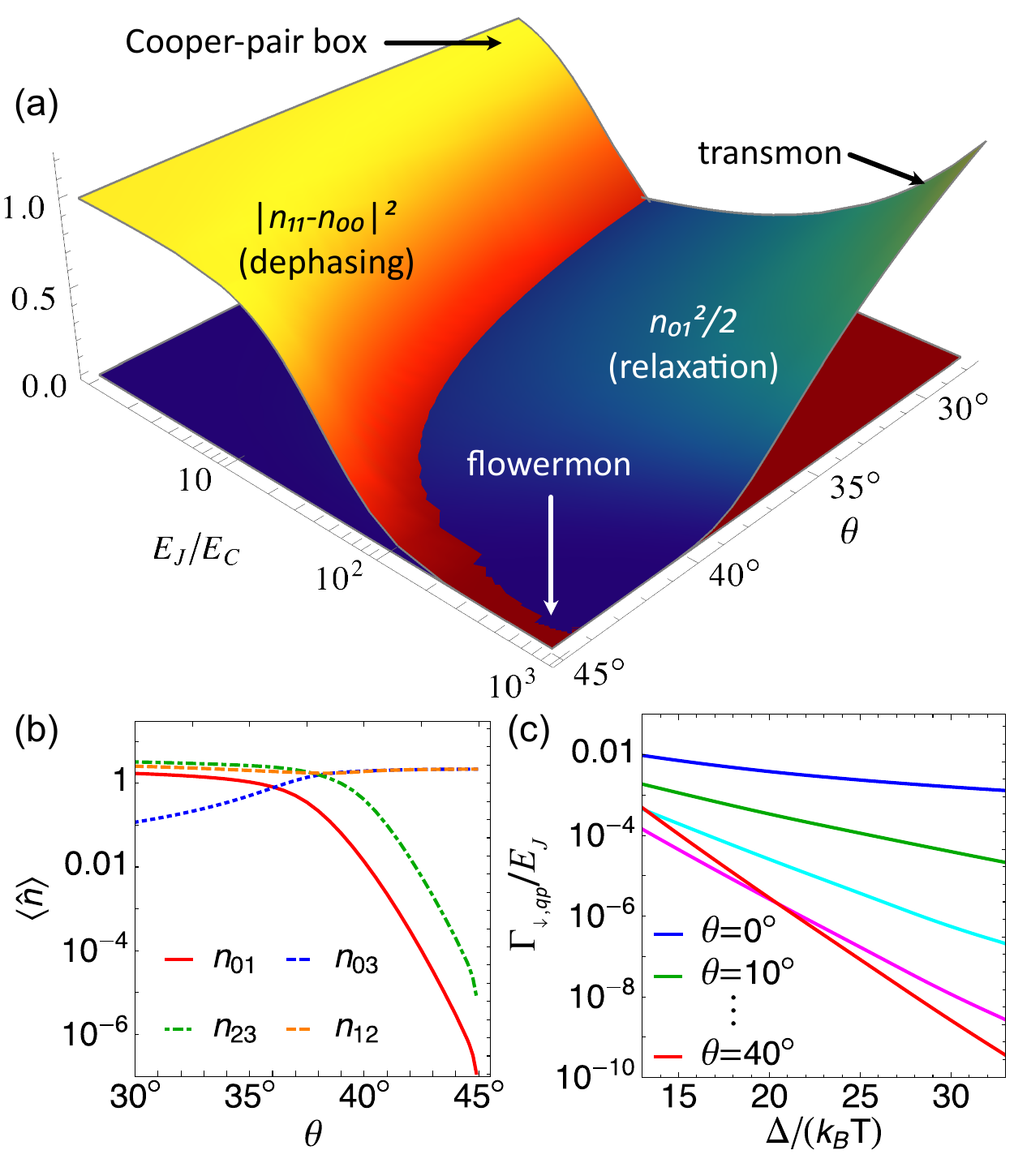}
		\caption{\textbf{Protection from decoherence}.\,\textbf{(a)}\,The charge matrix elements  $n_{01}^2$ and $|n_{11}-n_{00}|^2$, related to charge/capacitive relaxation and dephasing rates respectively, plotted vs. $E_{\mathrm J}/E_{\mathrm C}$ and the twisting angle $\theta$. Different regimes are shown, with the flowermon exhibiting protection from both relaxation and dephasing processes. \textbf{(b)} Charge matrix elements $n_{01}$, $n_{03}$, $n_{23}$, and $n_{12}$ vs. the twisting angle $\theta$ and $E_{\mathrm J}/E_{\mathrm C}=2000$. Both $n_{01}$ and $n_{23}$  decay exponentially as $\theta$$\rightarrow$$45^\circ$ but with a significant offset. \textbf{(c)} Dependence of the quasiparticle relaxation rate on the ratio $\Delta/k_{\mathrm B}T$ for different values of the\,twisting\,angles\,ranging\,between $\theta=0^\circ$ and $\theta=40^\circ$.}\label{fig-me}.
	\end{center}
\end{figure}

To illustrate the relevance of the flowermon symmetry properties for qubit protection, we consider the expression of the relaxation rate induced by the capacitive losses\,\cite{makhlin2001} 
\be \label{gamma1c}
\Gamma_{1c}= \frac{(8 E_{\mathrm C})^2}{\hbar^2}S_{n_g}(\omega_{01}) n_{01}^2\,, 
\ee
where $n_{xy}= \la\psi_x|\hat n |\psi_y\ra$ is the matrix element of the charge operator and $S_{n_g}(\omega_{01})$ is the spectral density of the capacitive noise at the qubit operational frequency. 
Due to the very small overlap in the charge basis $\hat{n}$ between $|\psi_0\ra$ and $|\psi_1\ra$ in the flowermon limit, the corresponding matrix element is exponentially  suppressed, preventing energy relaxation, see Fig.\,\ref{fig-me}(b). Looking at twisting angles $\theta\gtrsim40^\circ$, we, thus, expect an enhancement of the relaxation time over several orders of magnitude as compared to the standard transmon devices which are typically limited by such dielectric losses\,\cite{wang_surface_2015,ahmad2022}.


The corresponding dephasing rate due to charge fluctuations can be estimated as
\be
\Gamma_{\mathrm{\vf}c}\simeq \frac{32 E_C^2}{\hbar^2} S_{n_g}(0)|n_{11}-n_{00}|^2\,,
\ee 
where $S_{n_g}(0)$ is the spectral density of the charge noise at zero frequency. In the case where $E_{\mathrm C}$ is sufficiently small with respect to $E_{\mathrm J}$ and  $E_{\mathrm{\kappa}}$, the coefficient $n_{11}-n_{00}$ is exponentially suppressed. The suppression factor can be quantitatively estimated by mapping the Hamiltonian onto a tunneling problem, see Supplementary Information \,\cite{supplementary}.
Thanks to the large shunt capacitor, the flowermon inherits protection against dephasing induced by the charge noise from the transmon.

In superconducting qubits based on the $s$-wave junctions, the existence of an energy gap guarantees that the decoherence induced by thermally-equilibrium quasiparticles is exponentially suppressed\,\cite{catelani2011}. 
Here we show that for sufficiently clean interfaces, an analogous gap behavior is also present in the flowermon at high twisting angles.
To that end, we write the junction Hamiltonian as\,\cite{can2021} 
\be H=H_{\mathrm L}+H_{\mathrm R}+H_{\mathrm T}\,,
\ee
where $H_{\mathrm L}$ and $H_{\mathrm R}$ represent the BCS Hamiltonians of the two layers, i.e.,
\begin{eqnarray}
H_{\mathrm{L,R}}=\sum_{{\bf k}\sigma} \xi_{{\bf k}{\mathrm{L,R}}}c^\dag_{{\bf k}\sigma \mathrm{L,R}}c_{{\bf k}\sigma {\mathrm{L,R}}}+\nonumber \\
+\sum_{\bf k}\lf(\Delta_{{\bf k}\mathrm{L,R}}c^\dag_{{\bf k}\uparrow \mathrm{L,R}}c^\dag_{{\bf -k}\downarrow \mathrm{L,R}}+{\rm H.c.}\rg)\,,
\end{eqnarray}
with $c^\dag_{{\bf k}\sigma \mathrm{L,R}}$ and  $c_{{\bf k}\sigma \mathrm{L,R}}$ denoting creation and annihilation operators of electrons with spin $\sigma$ and momentum ${\bf k}$ in layers $L$ and $R$, while $H_{\mathrm T}$ represents the interlayer tunneling 
\be
H_{\mathrm T}=\sum_{{\bf kp}\sigma}\lf(t_{\bf kp}e^{-i\vf/2}c^\dag_{{\bf k}\sigma {\mathrm L}}c_{{\bf p}\sigma {\mathrm R}}+{\rm H.c.}\rg).
\ee
The  Hamiltonian $H$ thus depends on a small number of parameters, namely, the $d$-wave gap functions defined as 
\be\Delta_{{\bf k}{\mathrm L}}=\Delta_d \cos(2\theta_{\bf k})\,\,{\rm and}\,\,  \Delta_{{\bf k}{\mathrm R}}=\Delta_d \cos(2\theta_{\bf k}-2\theta),
\ee
with  $\theta_{\bf k}$ denoting the polar angle in the plane $k_x,k_y$,
the electronic dispersion around the Fermi surface $\xi_{\bf k}=\hbar^2 k^2/(2m)-\mu$, and the interlayer tunnel amplitude $t_{\bf kp}$.
The latter plays a central role in determining the junction's properties, as discussed in Ref.\,\cite{haenel2022}.
Following Refs.\,\cite{yokoyama2007,can2021,bruder1995}, we consider  the coherent tunneling limit with the in-plane momentum conservation, $t_{\bf kp}=t\,\delta_{\bf k_{\parallel}p_{\parallel}}$, suitable to describe weakly disordered $c$-axis junctions \cite{bruder1995}.
More accurate microscopic treatment of  $t_{\bf kp}$, including momentum dependence, has been previously considered\,\cite{Barash1995} and can affect the Josephson coupling terms themselves in addition to quasiparticle tunneling \,\cite{song_doping_2022}. This full treatment is beyond the scope of this work and will be the subject of a forthcoming publication.

We calculate the quasiparticle-induced relaxation rate using the approach developed by Refs.\,\cite{catelani2011,catelani2012,catelani2014}, which can be expressed as\,\cite{supplementary}
\be\label{qprate}
\Gamma_{\downarrow,{\rm qp}}=t^2|\la\psi_0|\sin(\vf/2)|\psi_1\ra|^2 S^{\rm \theta}_{\rm qp}(\omega_{01}),
\ee 
where $S^{\rm \theta}_{\rm qp}(\omega_{01})$ is the spectral density of the noise fluctuations induced by quasiparticle tunneling. 
In the above equation, the $\theta$ dependence of the relaxation rate can be traced back to the dependence of both the matrix element $\la\psi_0|\sin(\vf/2)|\psi_1\ra$ and spectral density on the twisting angle. 

\begin{figure}[t]
	\begin{center}
		\includegraphics[width=\columnwidth]{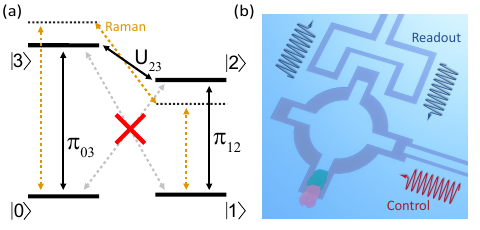}
		\caption{\textbf{Manipulation of the flowermon logical states}. \,\textbf{(a)}\, Logical states manipulation can be performed via the 2-nd and 3-rd excited states in a regime where the 0-1 matrix element and frequency are near 0, but the amplitude of the 2$\leftrightarrow$3 transition is still finite. The control is performed by standard microwave $\pi$-pulses between the 1-2, and 0-3 states, the desired control gate $U$ pulse between the 2-3 states, and again $\pi$-pulses between the 1-2, and 0-3 states. Alternatively, the 0-1 transition can be driven indirectly by a Raman process via simultaneous off-resonant driving of the 0-3, 2-3, and 2-1 transitions (orange arrows). Note that this protocol works even when $\omega_{01}\rightarrow 0$ because the selection rule forbids 0-2 and 1-3 transitions. \,\textbf{(b)}\, A sketch of a possible implementation of the device with control and readout lines.}\label{fig-control}.
	\end{center}
\end{figure}

Figure\,\ref{fig-me}(c) shows the results of the numerical calculation of the quasiparticle relaxation rate given by Eq. \eqref{qprate}. 
At high temperatures, high twisting angles correspond to a larger quasiparticle relaxation rate due to the larger prefactor $\la\psi_0|\sin(\vf/2)|\psi_1\ra$.  
On the contrary, in the low-temperature limit, high twisting angles  yield a strong exponential suppression of the quasi-particle rate $\Gamma_{\downarrow,{\rm qp}}$, as nodal quasiparticles are forbidden from tunneling due to a momentum mismatch.
The numerical results are well fitted by the simple analytical formula, $S^{\rm \theta}_{\rm qp}(\omega_{01})\propto\,e^{-\Delta_d/(k_BT) \sin(2\theta)^2}$, where, as mentioned above, $\Delta_d$ is the $d$-wave gap in the quasiparticle spectrum at zero twisting angle; this result is reminiscent of what was found in Ref.\,\cite{yokoyama2007}, see supplementary material\,\cite{supplementary} for details. 
This fit indicates that under appropriate conditions the quasi-particles in twisted cuprate vdW heterostructures are effectively gapped, in similarity to what happens in the s-wave junctions.

 So far we considered only thermal quasiparticles but superconducting devices are often limited by non-equilibrium quasiparticles\,\cite{martinis_energy_2009,de_visser_evidence_2014,vool_non-poissonian_2014,wang_measurement_2014,serniak_hot_2018,vepsalainen_impact_2020,cardani_reducing_2021}. 
We expect that the gap behavior highlighted here will similarly protect the flowermon from the decoherence induced by non-equilibrium quasiparticle tunneling\,\cite{Pop2014}. 
Note that similar protection can be obtained by an s/d-wave Josephson junction\,\cite{Ioffe1999,Patel2023}.

The decoupling of the flowermon qubit from noise sources hampers the direct control and readout of the qubit since the qubit energy $\hbar \omega_{01}$ (see Fig.\,\ref{fig2}(a)) and the charge matrix element $n_{01}$ (Fig.\,\ref{fig-me}(b)) are exponentially suppressed as $\theta$$\rightarrow$$45^\circ$. 
However, the structure of the flowermon spectrum allows for manipulation through the higher energy levels without sacrificing protection, see Fig.\,\ref{fig-control}(a). 
Specifically, there is a wide range of angles where the qubit matrix element $n_{01}$ is suppressed while the coupling between the 2-nd and 3-rd excited states $n_{23}$ is still finite (Fig.\,\ref{fig-me}b), e.g., $\theta=40^\circ$ corresponds to $n_{01}=0.01$ and $n_{23}=0.4$.
Thus, one can use the excited states for manipulation. For example, a $\pi/2$-pulse in the logical 0-1 space can be performed using standard microwave $\pi$-pulses between the 1-2, and 0-3 states, a $\pi/2$-pulse between the 2-3 states, and again $\pi$-pulses between the 1-2, and 0-3 states. 
To avoid populating the higher states, 0-1 manipulation can also be performed via a simultaneous application of the off-resonant Raman drives through the same energy levels\,\cite{vool2018driving}, see the orange line in Fig.\,\ref{fig-control}(a). 
Note that this scheme works even if $\omega_{01}$ is smaller than the frequency selectivity of the control pulses since the transitions 0-2, 1-3 are forbidden by the selection rules $n_{02}=n_{13}=0$. The higher excited states can be used similarly for measuring the state of the qubit, either by measuring non-protected transitions such as 0-3, 1-2 directly\,\cite{cottet2021electron} or through the dispersive coupling of these transitions to a cavity mode\,\cite{smith2020}. Figure\,\ref{fig-control}(b) shows a possible experimental implementation of such a qubit with the control and readout lines. A more quantitative discussion of this design with realistic qubit parameters is given in\,\cite{supplementary}.

In conclusion, we have developed a qubit design based on a twisted cuprate heterostructure, which we call the flowermon. The simple design of the flowermon, a single junction shunted by a large capacitance similar to that of a transmon, does not require flux tuning or precise control of the fabrication parameters. However, the $d$-wave structure of the superconducting phase in the unique junction endows the qubit with physical protection against decoherence. We have discussed the low energy spectrum structure of the qubit at various twist angles and quantitatively analyzed its sensitivity to decoherence mechanisms. Finally, we have presented schemes for the manipulation and readout of this qubit. At twisting angles close to $45^\circ$, the flowermon shows exponential suppression of the charge-induced noise, as well as protection from nodal quasiparticle tunneling. The flowermon is thus expected to provide the orders of magnitude improvement in the coherence time, promising remarkable progress in future superconducting quantum hardware. Furthermore, the protection offered by the qubit is directly related to the microscopic properties of the cuprate superconductor. Thus, the flowermon promises to become the prototype for a new class of hybrid devices which combine the benefits of quantum materials and coherent quantum circuits.

\textit{Acknowledgements--}
The work is partially supported by the Deutsche Forschungsgemeinschaft (DFG 452128813, DFG 512734967, DFG 492704387, and DFG 460444718) and co-funded by the European Union (ERC, cQEDscope, 101075962). The work of V.M.V. is supported by Terra Quantum Inc. and partially by the US NSF Grant Awards No.\,NSF 1809188 and 2105048. 
The authors are deeply grateful to Shu Yang Frank Zhao, Philip Kim, Bernard van Heck, Carlo Di Castro, Rosario Fazio, Andrei Zaikin, Kornelius Nielsch, Francesco Tafuri, Giampiero Pepe, Domenico Montemurro, Davide Massarotti, and Pavel A. Volkov for illuminating discussions.

\newpage
\setcounter{secnumdepth}{2}

\onecolumngrid
\appendix
\begin{center}
\large{\textbf{Supplementary Material for \\ ``A superconducting qubit based on cuprate twisted van der Waals heterostructure''}}
\end{center}

\section{Flowermon's spectrum}
\begin{figure}[b]
	\begin{center}
		\includegraphics[width=0.7\columnwidth]{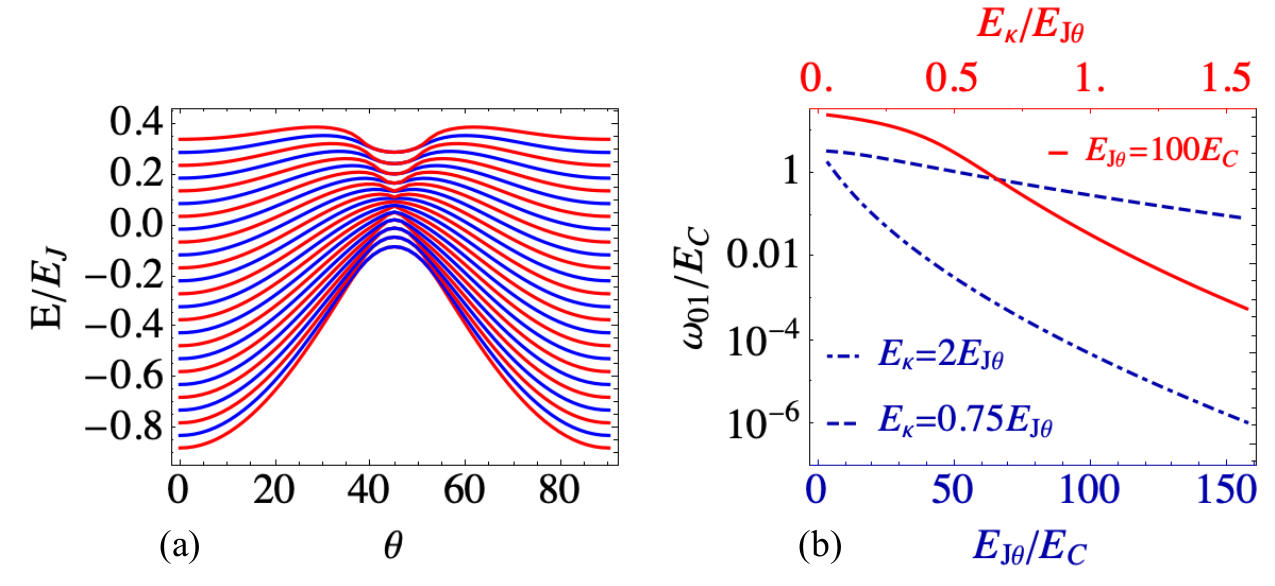}
		\caption{(a)Doublet structure  of the flowermon's  spectrum as a function of the twisting angle $\theta$ for $E_J=2000 E_C$ and $E_{\kappa}=0.1E_J$ and $n_g=0$. (b) Log plot showing the behavior of the lowest doublet energy splitting $\omega_{01}$. The red line shows the dependence  of $\omega_{01}$ on $E_{\kappa}/E_{C}$  for a fixed value of $E_{J\theta}=100\, E_C$ while the blue dashed and dot-dashed lines show the dependence of  $\omega_{01}$, on the ratio $E_{J\theta}/E_C$ for fixed values of $E_{\kappa}/E_{J\theta}=0.75,\,2$.}\label{fig-supp1}.
	\end{center}
\end{figure}

The flowermon's Hamiltonian, given by Eq.(1) of main text, can be recast as follows in the charge basis:
\be
H=4 E_C \sum_n (n-n_g)^2 |n\ra\la n|-\sum_n \lf( E_{J\theta}  |n\ra\la n+1|-E_{\kappa} |n\ra\la n+2|+{\rm H.c.}\rg)
\ee
where $|n\ra$ indicates a state of $n$ Cooper pairs.
Starting from the above Hamiltonian we calculate numerically the flowermon spectrum as a function of the different parameters governing the Hamiltonian, namely,   the twisting angle $\theta$, the gate charge $n_g$, the single- and double- Cooper pair tunneling amplitudes, $E_{J\theta}=E_J \cos \theta$ and $E_\kappa$, and the charging energy $E_C$. 
When $\theta=45^\circ$, the states with odd and even numbers of Cooper pairs are completely decoupled and the spectrum reduces to a set of degenerate doublets, as shown in Fig.\ref{fig-supp1}(a). 
For the low energy levels the degeneracy can be also related to the existence of degenerate classical minima at $\phi=\pm \phi_0=\pm\arccos E_{J\theta}/(4E_{\kappa})$. Indeed as  hinted at by  Fig.\ref{fig-supp1}(b) the splitting the lowest levels scales exponentially as a function of ratio $\Delta U_J/E_C$ where $\Delta U_J$ is the potential barrier height defined by Eq. (2) of main text.

\begin{figure}[t]
	\begin{center}
		\includegraphics[width=\columnwidth]{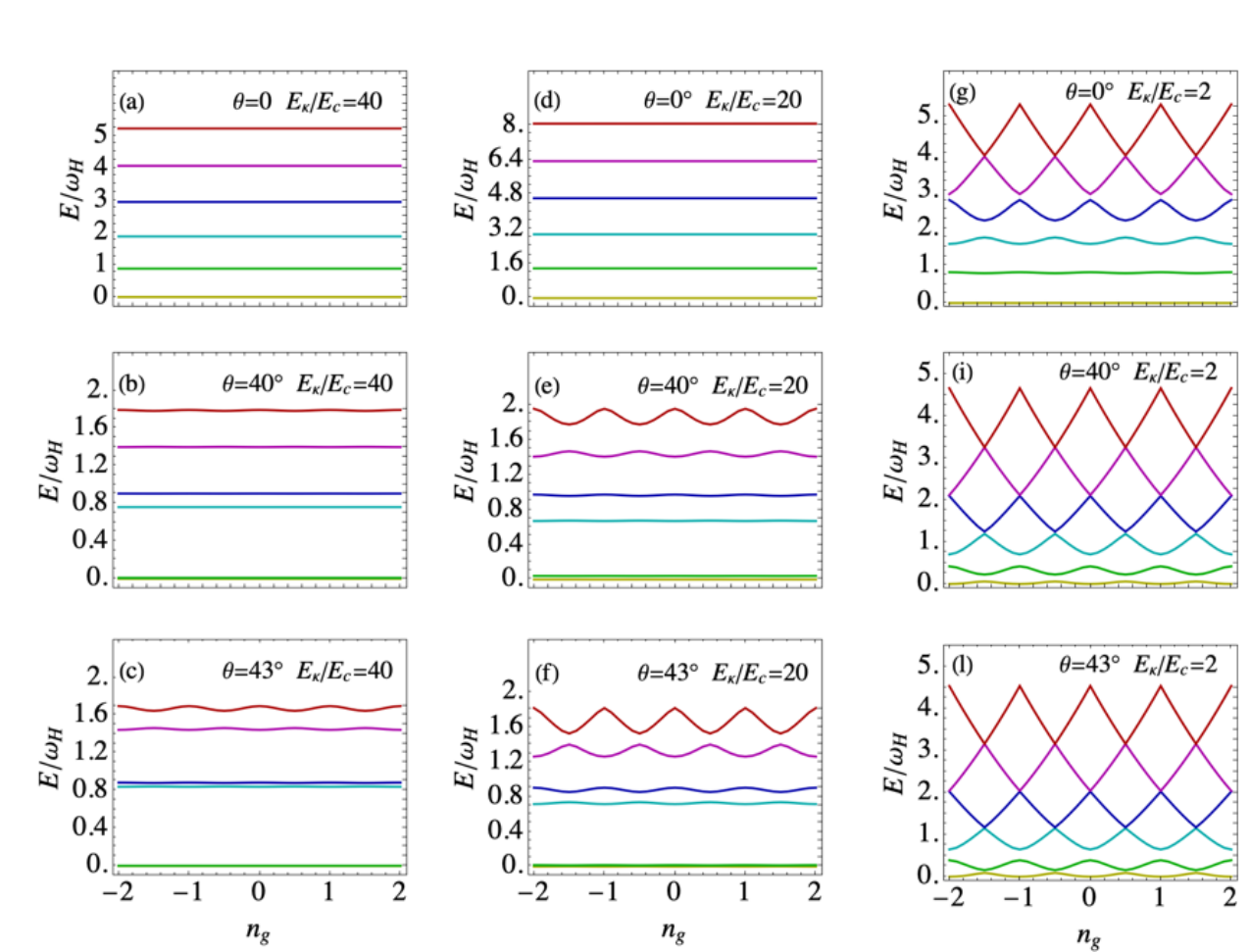}
		\caption{Structure of the low-energy spectrum of the flowermon as a function of $n_g$ for different values of $E_{\kappa}/E_C$ and different twisting angles.}\label{fig-sup2}.
	\end{center}
\end{figure}

In the phase basis the time-independent Schr\"odinger equation
%
has the form of a   Whittaker-Hill equation\cite{urwin1970}
\be
\frac{\partial^2\Psi_n}{\partial \varphi^2}+\lf[a \cos \vf-b\cos 2\vf \rg]\Psi_n=\lambda_n \Psi_n.
\ee
where we introduced the normalized parameters $a=E_{J\theta}/4 E_C$ and $b=E_{\kappa}/4 E_C$ and  the normalized eigenvalue $\lambda_n=E_n/4E_C$.
Analogously to the transmon, thanks to the large shunt capacitor, the flowermon is insensitive to gate charge fluctuations. Specifically, it has a spectrum that becomes flat as a function of $n_g$ both in the limit $E_J\gg E_\kappa\gg E_C$ (transmon) and in the limit  $E_\kappa> E_J\gg E_C$ (flowermon) as shown in Fig. \ref{fig-sup2}. In the two limits, we find an exponential decrease of the charge dispersion as we decrease $E_C$.
\onecolumngrid
\section{Quasiparticle induced decoherence}
Following the route outlined by Catelani et al.\cite{catelani2012}, starting from the microscopic Hamiltonian introduced in Ref. [\onlinecite{haenel2022,can2021}] and Eqs.(5-7) of main text, we cast the quasiparticle tunneling Hamiltonian as 
\be
H_T=\sum_{{\bf k p}\sigma}t_{\bf kp}\!\lf[\lf(u_{{\bf k}L}u_{{\bf p}R}-v_{{\bf k}L}v_{{\bf p}R}\rg)\cos \frac{\vf}{2}\gamma^\dag_{{\bf k}\sigma L}\gamma_{{\bf p}\sigma R}+i\lf(u_{{\bf k}L}u_{{\bf p}R}+v_{{\bf k}L}v_{{\bf p}R}\rg)\sin \frac{\vf}{2}\gamma^\dag_{{\bf k}\sigma L}\gamma_{{\bf p}\sigma R}{\rm +H.c.}\rg]\label{micro-tunn}
\ee
where the $\gamma$'s are creation and annihilation operators of Bogolyubov quasiparticles in the left and right lead, while $u_{\bf k M}$ and $v_{\bf k M}$ denote the Bogolyubov amplitudes of lead $M$
\be
u_{{\bf k}M}=\sqrt{1+\frac{\xi^2_{\bf k}}{E^2_{{\bf k}M}}}, \,\,v_{{\bf k}M}=\sqrt{1-\frac{\xi^2_{\bf k}}{E^2_{{\bf k}M}}}
\ee
where we assume $\xi_{\bf k}=\hbar^2 k^2/2m -\mu$ in both leads and  we  denote as usual as  $E_{{\bf k}M}$ the quasiparticle energies $E_{{\bf k}M}=\sqrt{\xi^2_{\bf k}+\Delta^2_{{\bf k }M}}$
 in lead $M$ with $\Delta_{{\bf k }M}$ indicating the corresponding gap-functions. The latter, denoting  as $\theta$ the twisting angle, can be expressed as
\be
\Delta_{{\bf k }L}=\Delta \cos(2\theta_{\bf k})\quad  \Delta_{{\bf k }R}=\Delta \cos(2\theta_{\bf k}-2\theta).
\ee
Projecting the Hamiltonian \eqref{micro-tunn} onto the qubit eigenstates basis and switching to the interaction picture for the quasi-particles we get:
\be\label{eq:V}
\hat V(t)=s_x N_x(t)\sigma_x+c_z N_z(t)\sigma_z
\ee
 where
 \bea
\!\!\!\!\!\!N_x(t)\!\!&=&\!\!\sum_{{\bf k p}\sigma}\lf[A^+_{{\bf kp}}i e^{i(E_{{\bf k}L}-E_{{\bf p}R})t}\tilde \gamma^\dag_{{\bf k}\sigma L}\tilde\gamma_{{\bf p}\sigma R}{\rm +H.c.}\rg]\!,\label{eq:Nx}\\ 
 \!\!\!\!\!\!N_z(t)\!\!&=&\!\!\sum_{{\bf k p}\sigma}\lf[A^-_{{\bf kp}}i e^{i(E_{{\bf k}L}-E_{{\bf p}R})t}\tilde \gamma^\dag_{{\bf k}\sigma L}\tilde\gamma_{{\bf p}\sigma R}{\rm +H.c.}\rg]\!,\label{eq:Nz} \\
 {\rm and } && s_x=\Tr\lf[\hat S \sigma_x\rg]\, {\rm and }  \, c_z=\Tr\lf[\hat C \sigma_z\rg] \eea
with the matrices $\hat S$ and $\hat C$ defined as  $\hat S_{ab}=\la \psi_a| \sin \phi/2|\psi_b\ra$ and  $\hat C_{ab}=\la \psi_a| \cos \phi/2|\psi_b\ra$ and 
\be
A^\pm_{{\bf kp}}=t_{\bf kp}\lf(u_{{\bf k}L}u_{{\bf p}R}\pm v_{{\bf k}L}v_{{\bf p}R}\rg).\label{eq:A}
\ee
Note that in deriving Eq.\eqref{eq:V} we used the approximate $\vf\rightarrow-\vf$ symmetry of the flowermon eigenstates, the symmetry becomes exact for  half-integer values of  $n_g$. 

Starting from Eq. \eqref{eq:V} we can calculate the quasiparticles-induced qubit's relaxation rate by means of Bloch-Redfield theory.  By doing so we obtain: 
\be
\Gamma_{\downarrow,{\rm qp}}=\frac{s_x^2}{\hbar^2}\int_{-\infty}^\infty \!\!\! \la N_x(t)N_x(0)\ra e^{-i\omega_{01}t} dt
\ee
that using Eqs.\eqref{eq:Nx} and \eqref{eq:A}  can be recast as  
\bea\label{gammasum}
\Gamma_{\downarrow,{\rm qp}}\!\!&=&\!\!\frac{s_x^2}{\hbar^2}\sum_{{\bf k p}}\!\int_{-\infty}^\infty \! \! t^2_{\bf kp} \lf(1+\frac{|\Delta_{{\bf k}L}\Delta_{{\bf p}R}|}{E_{{\bf k}L}E_{{\bf p}R}}\rg)f_{{\bf k}L}(1-f_{{\bf p}R})\cdot\nn\\& & \cdot e^{i(E_{{\bf k}L}-E_{{\bf p}R}-\omega_{01})t} dt.
\eea

Following Ref.[\onlinecite{yokoyama2007,can2021}] we consider the case of coherent tunneling with in-plane momentum conservation, $t_{\bf kp}=t\,\delta_{\bf k_{\parallel}p_{\parallel}}$, suitable to describe weakly disordered $c$-axis junctions.
This assumptions can be justified as discussed in Ref. \cite{bruder1995} considering the spatial dependence of the tunneling matrix element across the barrier.
A simple calculations shows that, starting from Eq. \eqref{gammasum} and  introducing the quasiparticle density of states, $\Gamma_{\downarrow,{\rm qp}}$
can be expressed  as
\be
\Gamma^{\rm \theta}_{\downarrow,{\rm qp}}=\frac{4 t^2}{\hbar}|\la\psi_0|\sin(\vf/2)|\psi_1\ra|^2  S^{\rm \theta}_{\rm qp}(\omega_{01})
\ee
where we  indicated as $S^{\rm \theta}(\omega_{01})$ the quasiparticle spectrum for coherent tunneling,
\begin{figure}[t]
	\begin{center}
		\includegraphics[width=0.5\columnwidth]{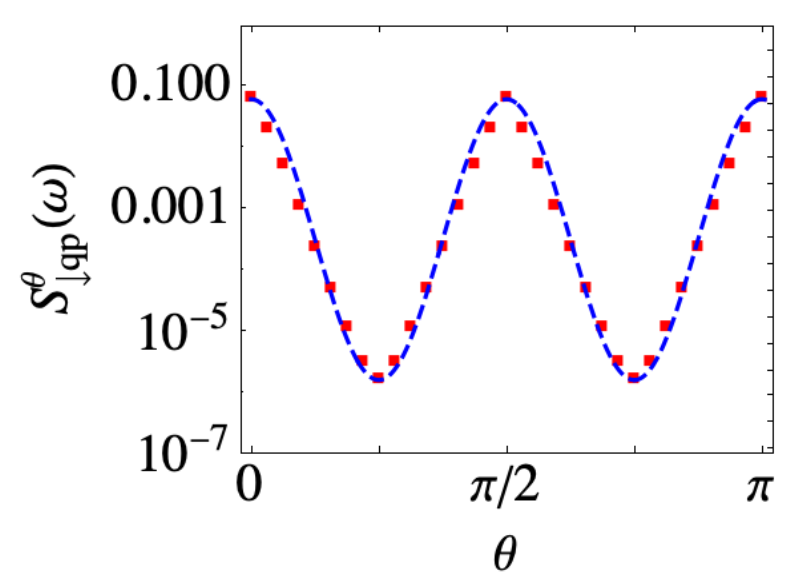}
		\caption{Dependence of the first two doublets splittings on the twisting angles. We set $E_J=1$THz, $E_c=500$MHz,$E_\kappa=0.1E_J$. Dependence of the quasiparticle spectrum on the twisting angle. The dot indicate the result of a numerical integration of Eq. \eqref{eq:gammaqpin} while the red dahsed line represent the fit to an exponential $e^\frac{-C \Delta \sin(2\theta)^2}{k_B T}$ with $C$ fitting coefficient.}\label{fig-sup}.
	\end{center}
\end{figure}
\begin{equation}\label{qp-spectrum}
S^{ \theta}_{\downarrow,{\rm qp}}(\omega_{01})=N_0^2\int \frac{d\theta_{\bf k}}{2\pi} \int_{\Delta_{{\bf k}L}} \!\!\!\!dE \int_{\Delta_{{\bf k}R}}\!\!\!\! dE^\prime \frac{E E^\prime+|\Delta_{{\bf k}L}\Delta_{{\bf k}R}|}{\sqrt{E^2-\Delta^2_{{\bf k}L}}\sqrt{{E^\prime}^2-\Delta^2_{{\bf k}R}}}f(E)(1-f(E^\prime))\delta(E-E^\prime+\omega_{01}).
\end{equation}
where $N_0$ denotes the density of states at the Fermi level.
It is useful to compare the above results with case of fully incoherent tunneling, $t_{\bf kp}=t$, applicable in the presence of strong disorder. In this case  we can write the relaxation rate as:
\be
\Gamma^{\rm in}_{\downarrow,{\rm qp}}=\frac{4 t^2}{\hbar}|\la\psi_0|\sin(\vf/2)|\psi_1\ra|^2 S^{\rm in}_{\rm qp}(\omega_{01})
\ee
where we  indicated as $S^{\rm in}_{\rm qp}(\omega_{01})$ the quasiparticle spectrum in the case of incoherent tunneling
\begin{equation}
S^{\rm in}_{\rm qp}(\omega_{01})=N_0^2\int \frac{d\theta_{\bf k}d\theta_{\bf p}}{4\pi^2} \int_{\Delta_{{\bf k}L}} \!\!\!\!dE \int_{\Delta_{{\bf p}R}}\!\!\!\! dE^\prime \frac{E E^\prime+|\Delta_{{\bf k}L}\Delta_{{\bf p}R}|}{\sqrt{E^2-\Delta^2_{{\bf k}L}}\sqrt{{E^\prime}^2-\Delta^2_{{\bf p}R}}}f(E)(1-f(E^\prime))\delta(E-E^\prime+\omega_{01}).\label{eq:gammaqpin}
\end{equation}
%
As expected the above equation yields a quasi-particle spectrum independent of the twisting angle. Furthermore, for a $s$-wave gap, {\sl i.e.} setting $\Delta_{{\bf k}L}=\Delta_s$, it reduces to the result of Catelani et al.\cite{catelani2011} {\sl i.e.}
\begin{equation}\label{gammaswave}
\Gamma^{\rm in}_{\downarrow,{\rm qp}} \rightarrow \frac{16 E_{J0}^s}{\hbar \,\pi} s_x^2 e^{-\Delta_s/(k_B T)}e^{-\omega/(2 k_B T)}K_0\lf(\frac{\omega}{2k_B T}\rg)
\end{equation}
where $K_0$ is the modified Bessel function of the second kind and we used $E_{J0}^s=N_0^2t^2\Delta_s/4\pi$.
 A more detailed microscopic analysis of the coherent and incoherent tunneling models and their relation with the properties of disorder may be found in Ref.[\onlinecite{haenel2022}].
Starting from Eq.\eqref{qp-spectrum}, we can calculate the quasiparticle-induced relaxation rate as a function of the quasiparticle gap $\Delta$ and the twisting angle. By doing so we obtain the results shown in Fig. 4 of main text. Furthermore, by fitting the numerical data, we were able to extrapolate a simple analytical behavior for the quasiparticle spectrum,  $\Gamma^{\rm \theta}_{\rm qp}(\omega_{01})\ \sim e^{-\Delta_d/(k_BT) \sin(2\theta)^2}$, reminiscent of the Yokoyama et al. result of Ref.\,\cite{yokoyama2007} as shown in Fig. \ref{fig-sup}.


\onecolumngrid

\onecolumngrid

\section{Design and qubit parameters}

The calculations in the manuscript were done using the energy ratio $E_J/E_c = 2000$. In this section, we will use realistic device parameters to obtain this regime, remembering that $E_c = {e^2}/{(2C)}$ and $E_J = \Phi_0 I_c/2\pi$ where $C$ is the total capacitance and $I_c$ is the critical current. As a benchmark for operation at typical GHz frequencies, we choose a Josephson energy $E_J$ of 1 THz and so the corresponding critical current is $I_c = 2$ $\mu A$. In particular, a BSCCO junction has a critical current density $J_c$ of 0.1 $kA/cm^2$ and then the area of the junction, i.e. the area where the two flakes are overlapped, should be of 2$\mu m^2$. BSCCO junctions with this area are not impossible to realize but they are challenging because the stacking process of the two flakes must be quick otherwise the interface will lose its superconducting properties, becoming an insulator. Typical current junction sizes are $\approx 1000\,\mu m^2$ and reducing the junction size is expected to be the most significant limitation to the implementation of the flowermon qubit.

\begin{figure}[h!]
\vspace{-0.5cm}
	\begin{center}
		\includegraphics[width=\columnwidth]{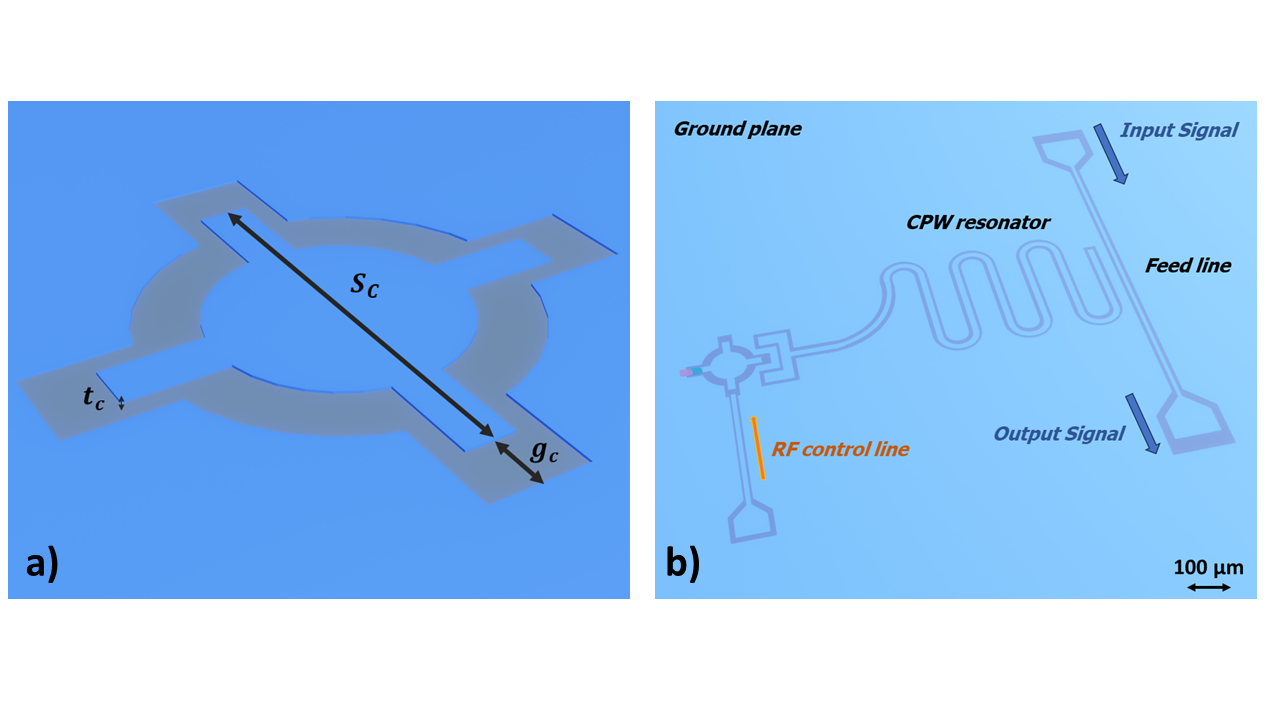}
        \caption{\textbf{Control circuit}. a) Parallel capacitor of $40fF$, the geometrical dimensions of the capacitor are displayed in the figure. The capacitor thickness ($t_c$) is 50 nm, the distance between the two plates of the capacitor ($g_c$) is 40 $\mu m$ and the total length of the central plate ($S_c$) is 180 $\mu m$. b) The capacitor is coupled to a CPW $\lambda/2$ resonator with the following geometrical parameters: width of 12 $\mu m$, a gap of 1 $\mu m$ and a length of 7 mm so that its resonance frequency is around 8.4 GHz.
        The signal is transmitted to the resonator through a transmission line (Feed line) matched to 50 $\Omega$ with a width of 5 $\mu m$ and a gap of 3 $\mu m$. }
		\label{supp_fig2}
          
	\end{center}
\end{figure}

To achieve the corresponding capacitive energy $E_c = 500$ MHz, so that $E_J/E_c = 2000$, we need a total capacitance $C$ of about $40fF$. This is well within achievable parameters, and using the simulation tool Ansys we designed and simulated a large parallel capacitor, adapting its shape to obtain the desired value of $C$. The resulting design is displayed in Figure \ref{supp_fig2}a. The capacitor thickness ($t_c$) is 50 nm, the distance between the two plates of the capacitor ($g_c$) is 40 $\mu m$ and the total length of the central plate ($S_c$) is 180 $\mu m$.

\vspace{0.3cm}

This parallel plate capacitor will be capacitively coupled with a Coplanar Waveguide (CPW) resonator to perform the readout and to a radio-frequency line to control the state of the qubit (Figure \ref{supp_fig2}b).

\end{document}